\documentclass[preprint,onecolumn,nofootinbib]{revtex4}
\usepackage[colorlinks=true,linkcolor=blue,urlcolor=blue,filecolor=black,citecolor=red,pdfstartview=FitV,pdftitle={},pdfsubject={},pdfkeywords={},pdfpagemode=None,bookmarksopen=true]{hyperref}
\usepackage{graphicx}
\usepackage{amsmath}
\usepackage{amsfonts}
\usepackage{amssymb,ulem}
\usepackage{color}%
\usepackage{tikz}
\usepackage{dcolumn}
\usepackage{xcolor}

\begin{document}
\title{
  Entanglement Wedge Minimum Cross-Section for Holographic Aether Gravity
}
\author{Chong-Ye Chen $^{1}$}
\email{cycheng@stu2018.jnu.edu.cn}
\author{Wei Xiong $^{1}$}
\email{phyxw@stu2019.jnu.edu.cn}
\author{Chao Niu $^{1}$}
\email{niuchaophy@gmail.com}
\author{Cheng-Yong Zhang $^{1}$}
\email{zhangcy@email.jnu.edu.cn}
\author{Peng Liu $^{1}$}
\email{phylp@email.jnu.edu.cn}
\thanks{corresponding author}
\affiliation{
  $^1$ Department of Physics and Siyuan Laboratory, Jinan University, Guangzhou 510632, China
}

\begin{abstract}

  We study the entanglement wedge cross-section (EWCS) in holographic Aether gravity theory, a gravity theory with Lorentz symmetry violation while keeping the general covariance intact. We find that only a limited parameter space is allowed to obtain a black brane with positive Hawking temperature. Subject to these allowed parameter regions, we find that the EWCS could exhibit non-monotonic behaviors with system parameters. Meanwhile, the holographic entanglement entropy (HEE), and the corresponding mutual information (MI), can only exhibit monotonic behaviors. These phenomena suggest that the EWCS could capture much more rich content of the entanglement than that of the HEE and the MI. The role of the Lorentz violation in determining the behaviors of quantum information-related quantities is also analyzed.

\end{abstract}

\maketitle

\tableofcontents

\section{Introduction}
\label{sec:introduction}

Quantum information, as a unique feature of quantum systems, is gaining increasing attention from areas such as holographic duality theory and condensed matter theory. Many measures have been adopted to study quantum entanglement, including entanglement entropy (EE), mutual information (MI), R\'enyi entropy, and so on. According to their definitions, they can describe different degrees of freedom of quantum systems. Among them, EE is the most well-known physical quantity to capture the entanglement of a pure state. However, EE is not suitable for describing the entanglement of mixed states. Some other entanglement-related quantities, such as the MI, the R\'enyi entanglement entropy, entanglement of purification (EoP), and reflected entropy can be better candidate measures of mixed state entanglement \cite{vidal:2002,Horodecki:2009review}. However, one of the main problems with quantum entanglement is that it is notoriously difficult to calculate.

Recently, gauge/gravity duality has been widely used to study strongly correlated physics \cite{Ling:2015exa,Landsteiner:2019kxb,Ling:2014laa,Ling:2014bda}. In addition to its power in strongly correlated systems, such as condensed matter theory and QCD theory, gauge/gravity duality also associates the information-related physical quantities in strongly correlated systems with geometric quantities in dual gravitational systems \cite{Sakai:2004cn,Donos:2012js}. The entanglement entropy of the dual quantum field theory has been proposed proportional to the area of the minimum surface in the dual bulk geometry, which was dubbed as the holographic entanglement entropy (HEE) \cite{Ryu:2006bv}. After that, many other holographic duals of quantum information-related physical quantities have been proposed and studied \cite{Nishioka:2006gr,Klebanov:2007ws,Pakman:2008ui,Zhang:2016rcm,Zeng:2016fsb,Kudler-Flam:2020url,BabaeiVelni:2020wfl,Sahraei:2021wqn,Khoeini-Moghaddam:2020ymm,KumarBasak:2020eia}. For example, the R\'enyi entropy has been proposed proportional to the area of the minimum cosmic brane \cite{Dong:2016fnf}, which back-reacts on the background geometry. Meanwhile, the entanglement of purification, reflected entropy and the entanglement negativity have all been related to the minimum cross-section in the entanglement wedge (EWCS) \cite{Takayanagi:2017knl,Nguyen:2017yqw,Kudler-Flam:2018qjo,Kusuki:2019zsp,Dutta:2019gen,Gong:2020pse,Fu:2020oep,Liu:2021stu,Zhang:2021edm,Tamaoka:2018ned}. In addition, quantum complexity has been proposed proportional to the volume or the action of a certain region, that can capture quantum information other than the entanglement \cite{Susskind:2014rva}. Furthermore, the butterfly velocity that measures the speed of the propagation of quantum information has been found related to the horizon of the black hole \cite{Shenker:2013pqa,Sekino:2008he,Maldacena:2015waa,Donos:2012js,Blake:2016wvh,Blake:2016sud,Ling:2016ibq,Ling:2016wuy,Wu:2017mdl,Liu:2019npm}. All of the above progress become the cornerstone of the study of quantum information of the strongly related systems in the framework of the holographic duality.

The Lorentz invariance is one of the principles of general relativity. However, the violation of Lorentz invariance in the condensed matter systems is common. To study the condensed matter systems with more practical significance, the violation of Lorentz invariance in holographic duality theory has been widely studied and found to exist in gravitational systems such as massive gravity theories \cite{Vegh:2013sk,Blake:2013bqa,Blake:2013owa,Davison:2013jba,deRham:2014zqa,Baggioli:2014roa,Alberte:2015isw}. However, a more natural approach is to explicitly break the Lorentz invariance in the gravity theory. Recently, the theory of Aether gravity has been proposed and studied as a kind of gravity theory that breaks the Lorentz symmetry but preserves the general covariance \cite{Kai:2015nlbh,Kai:2017ea}. Though HEE has been widely studied in many different holographic theories \cite{Ling:2015dma,Ling:2016wyr}, the properties of mixed states entanglement such as MI and EWCS in many holographic theories are still unclear. The main purpose of this paper is to investigate the effect of Lorentz symmetry violation on the entanglement of mixed states by studying EWCS, MI, and HEE in the holographic Aether gravity theory.

In section \ref{sec:alg} we will introduce the AdS Aether gravity model. Then we discuss the properties of HEE (\ref{sec:HEE}), MI (\ref{sec:mi}) and EWCS (\ref{sec:eop_phenomena}). Lastly, we give a summary in \ref{sec:discuss}.

\section{Charged static solutions of Einstein-aether theory}\label{sec:alg}

\subsection{Introduction to holographic Aether Gravity}

The action of the $n$-dimensional Einstein-Aether-Maxwell theory reads \cite{Kai:2017ea},
\begin{equation}\label{eq:action2}
  S = \int d^n x \frac{\sqrt{-g}}{16\pi G_{ae}}(R - 2 \Lambda + \mathcal{L}_{ae}
  - \alpha F_{\mu \nu}F^{\mu \nu}),
\end{equation}
and
\begin{equation}\label{eq:Legendreae1}
  \begin{aligned}
    \mathcal{L}_{ae} = & c_1 (\nabla_{\mu} u_{\nu})(\nabla^{\mu} u ^{\nu})+ c_{2} (\nabla^{\nu} u _{\nu})^2 +c_{3} (\nabla_{\nu} u _{\mu})(\nabla^{\mu} u ^{\nu})                                         \\
                       & - c_{4} u^{\mu} u^{\nu} (\nabla_{\mu} u _{\rho})(\nabla_{\nu} u ^{\rho})
    +\lambda (u_{\nu} u^{\nu}+1).
  \end{aligned}
\end{equation}
$F = dA$, and $A$ is the Maxwell field. $G_{ae}$ in \eqref{eq:action2} is the constant related to Newton's gravitational constant $G_{N}$ by $G_{ae}=(1-c_{14}/2)G_N$, where $c_{ij}\equiv c_i+c_j$. The $c_1$, $c_2$, $c_3$ and $c_4$ the coupling constants, and $\lambda$ the Lagrange multiplier such that the aether vector $u_{a}$ satisfies the timelike constraint $u_{a} u^{a}=-1$. The aether vector $u_{a}$ can also be expressed as the one-form of a scalar field $\phi$,
\begin{equation}\label{eq:khrononfield}
  u _ { a } \equiv \frac { - \partial _ { a } \phi } { \sqrt { - g ^ { bc } \partial _ { b  } \phi \partial _ { c} \phi } },
\end{equation}
which is called the khronon scalar \cite{Kai:2017ea,Kai:2015nlbh}. The speed of the khronon scalar, defined as the speed of the mode of the perturbation of $\phi$, is given by
\begin{equation}\label{eq:cphi}
  c_{\phi}^2=\frac{c_{123}}{c_{14}}.
\end{equation}
Eq. \eqref{eq:cphi} shows that the most significant cases are $c_{123} = 0$ $(c_{\phi}^2 = 0)$ and $c_{14} = 0$ $(c_{\phi}^2 \to \infty)$. Specifically, we choose the case $c_{14} = 0 $ in this work.
The ansatz of solution is given by
\begin{equation}
  \label{aeansatz}
  \begin{aligned}
    ds^2    & =- F(r)dv^2 + 2 dr dv + r^2 h_{ij} dx^i dx^j \\
    u^a     & =u^v(r)(e_v)^a+V(r)(e_r)^a           \\
    A_a     & =A_0(r)(e_v)_a 
  \end{aligned}
\end{equation}
where $u^v(r)=(V(r)+\sqrt{F(r)+V(r)^2})/F(r)$. The form of $h_{ij}dx^i dx^j$ is given by
\begin{equation}
  \label{eq:hijform}
  h_{ij}dx^i dx^j = \begin{cases}d \theta^{2}+\sin ^{2} \theta d \Omega_{n-3}^{2} & k=1 \\ d x_{i} d x^{i} & k=0 \\ d \theta^{2}+\sinh ^{2} \theta d \Omega_{n-3}^{2} & k=-1\end{cases} \ i,j =1,2\ \cdots\ n-2.
\end{equation}
There are three options, sphere, planar or hyperbolic spacetime metric that can be used as induced metric, corresponding to $k=1,0,-1$. Specifically, we choose $k=0$, and induced metric component $h_{ij}$ takes the form of $\delta_{ij}$. In the case of $c_{14} = 0$, $n=4$ and $k=0$, we follow from \cite{Kai:2017ea} and the solution is given by
\begin{equation}
  \label{eq:fandv}
  \begin{aligned}
    A_0(r) & =A_c-\frac{Q}{r} \\
    F(r)   & = \frac{2 M_{a}}{r} + \frac{Q^2}{r^2}+\frac{4 c_{13} B^2}{r^4} -\frac{2\Lambda-\rho_a B^2 r_{z}^{-6}}{6}r^2 \\
    V(r)   & = 2B r(r_z^{-3}-r^{-3})
  \end{aligned}
\end{equation}
where $\rho_{a}= 24c_{13}-3c_{123}$. $A_c$, $B$ and $r_{z}$ are constants. $M_a$ and $Q$ are the mass and charge of the black hole respectively. $\Lambda$ is the cosmological constant.
For convenience, we let $r_z \to \infty$, $\beta \equiv c_{13} B^2$ and $\Lambda = -3$. The function $F$ in \eqref{eq:fandv} becomes
\begin{equation}
  \label{eq:fexps}
  \begin{aligned}
    F(r)= & \frac{2 M_{a}}{r} + \frac{{Q}^2}{r^2}+\frac{4 \beta}{r^4} +r^2.
  \end{aligned}
\end{equation}
For later calculation of the minimum surface, we transform the coordinate system and the metric takes the form
\begin{equation}
  \label{eq:aemetric}
  d s ^ { 2 } = - F ( r ) d t ^ { 2 } + F ^ { - 1 } ( r ) d r ^ { 2 } + r ^ { 2 } d x _ { i } d x ^ { i } , \quad i  = 1,2.
\end{equation}
At the horizon $r=r_h$ we have $F(r_h)=0$. We can see that $r_{h}$ determines $M_a$ and the Hawking temperature reads,
\begin{equation}
  \label{eq:hawkingtemp}
  T= \frac{F'(r_h)}{4\pi} = -\frac{12\beta + Q^2 r_{h}^2 -3 r_{h}^6 }{4 \pi r_{h}^5},
\end{equation}
The system can be specified by $(T,\beta,Q)$. In addition, the aether vector induces the Lorentz violation,
and from the aether action \eqref{eq:Legendreae1} we see that $c_1,\,c_2,\,c_3,\,c_4$ associate with it. However, for the solution that we are addressing, the overall effect of Lorentz violation is only reflected in the parameter $\beta$, which we name as the Lorentz violation parameter.

\subsection{Calculation of allowed parameter region}

\begin{figure}[]
  \centering
  \includegraphics[width =0.45\textwidth]{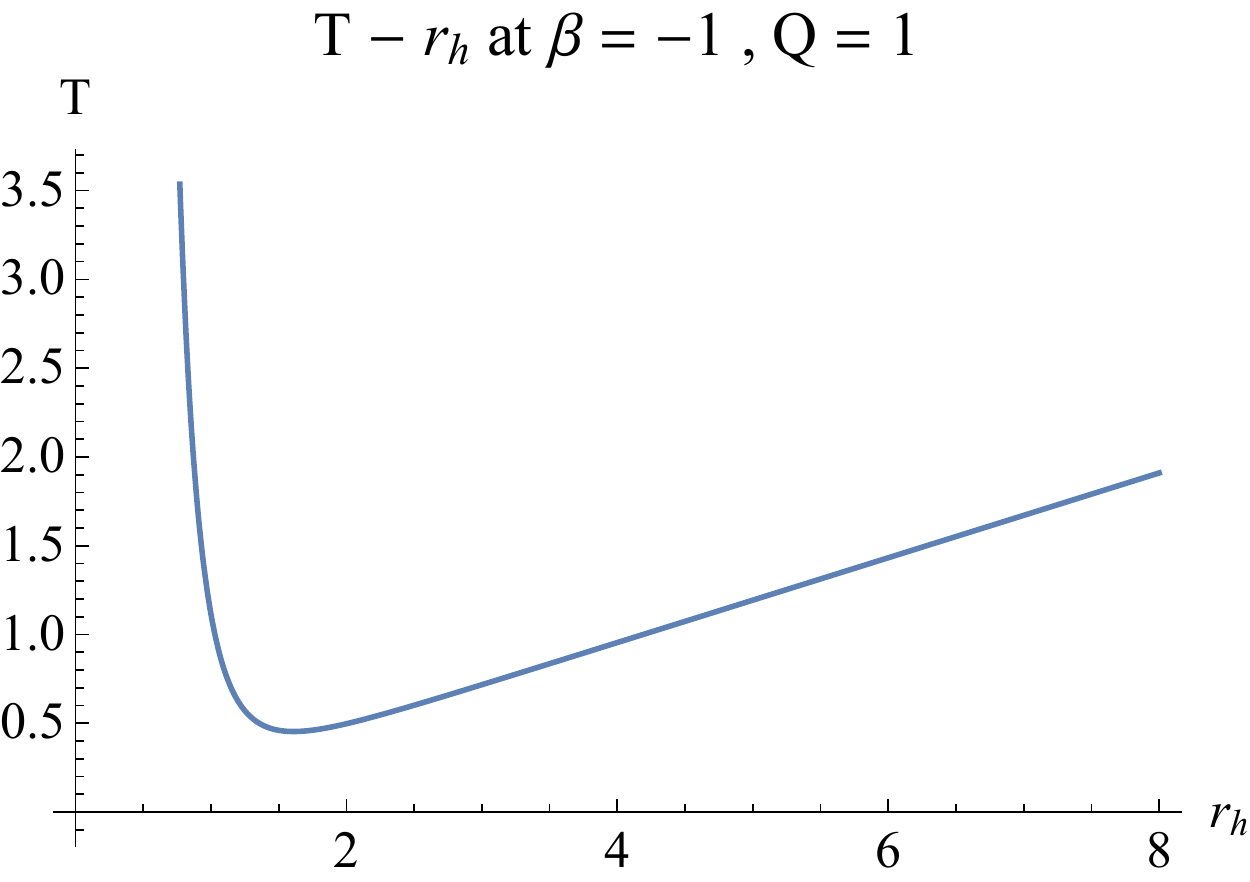}
  \includegraphics[width =0.45\textwidth]{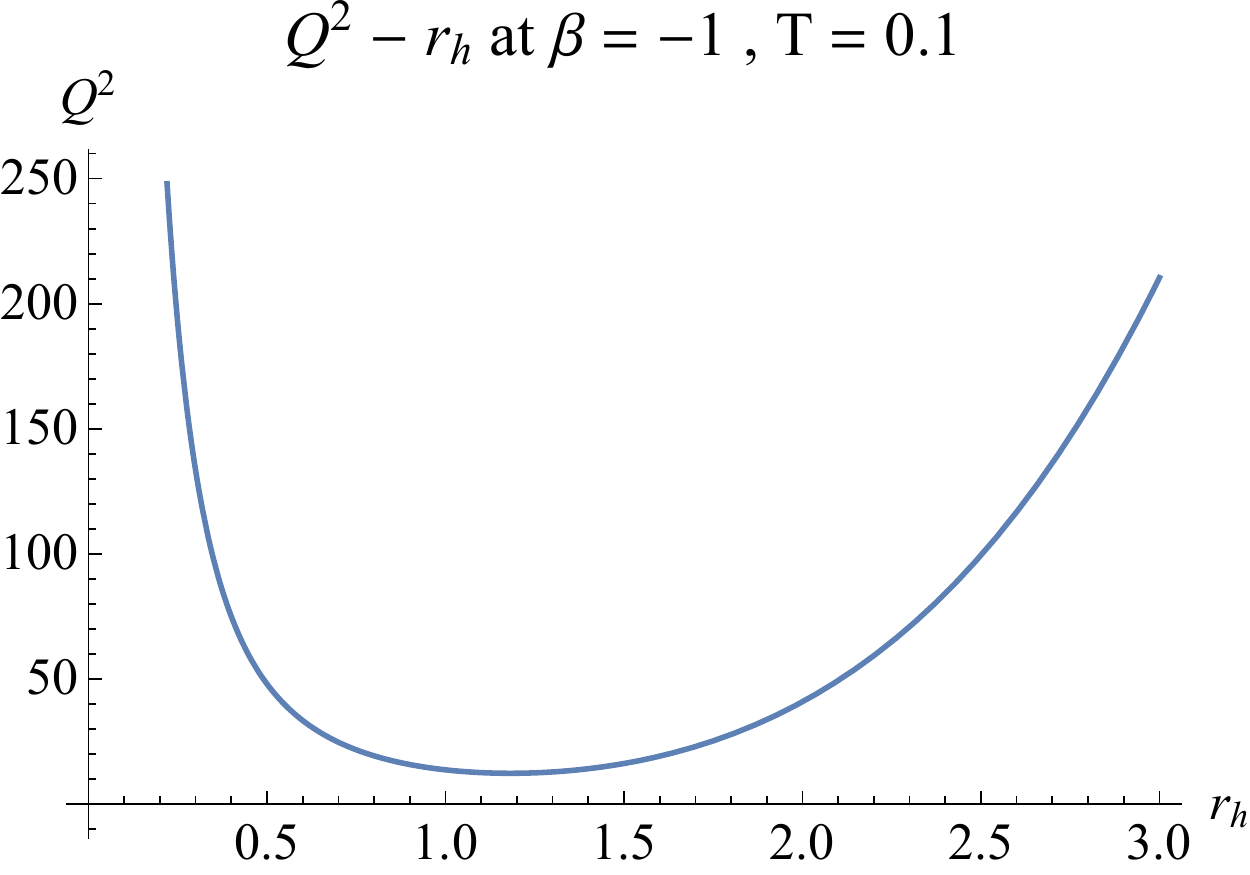}
  \caption{The left plot: $T$ vs $r_h$ with fixed $Q$ and $\beta$. Apparently, $T$ has a lower bound here. The right plot: $Q$ versus $r_h$ with fixed $T$ and $\beta$. $Q$ has a lower bound here.}
  \label{fig:openup}
\end{figure}

A reasonable black hole system should have a non-negative Hawking temperature $T$ and a non-negative horizon radius $r_h$. In addition, $Q$ always appears as a squared term, so we only discuss the case $Q \geqslant 0$ here. In summary, we discuss the physically reasonable parameter regions satisfying,
\begin{equation}\label{eq:reasonable}
  T\geqslant 0,\; r_h\geqslant 0,\; Q\geqslant 0.
\end{equation}
The Hawking temperature \eqref{eq:hawkingtemp} suggests that not any choice of $\beta$ satisfies \eqref{eq:reasonable}, hence we numerically work out the allowed regions. Firstly, for $\beta < 0$, the relationship between $T$ and $r_h$, $Q^2$ and $r_h$ are two non-monotonic functions, and $T$ and $Q$ can have lower bounds (see Fig. \ref{fig:openup}).
Then we can get the minimum values of $Q$ or $T$ by locating the extreme points. These extreme points are also called critical parameters, and the surface formed by them is the boundary between the allowed parameter regions and the non-physical parameter regions. For $\beta > 0$, however, the situation is different. It can be derived from the Hawking temperature \eqref{eq:hawkingtemp} that if the root $r_h$ is very large, the corresponding $T$ will also be very large; while when the root $r_h$ tends to $0$, the corresponding $T$ will be negative infinity. Next, we show the numerical results of the allowed parameter regions.

\begin{figure}[]
  \centering
  \includegraphics[width =0.32\textwidth]{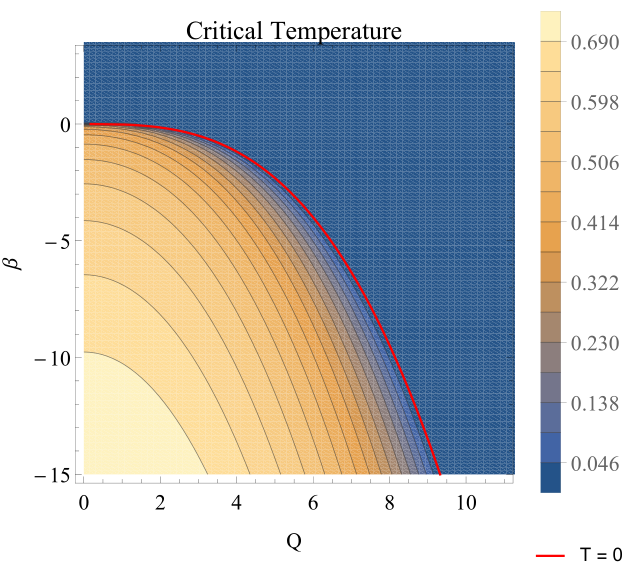}
  \includegraphics[width =0.32\textwidth]{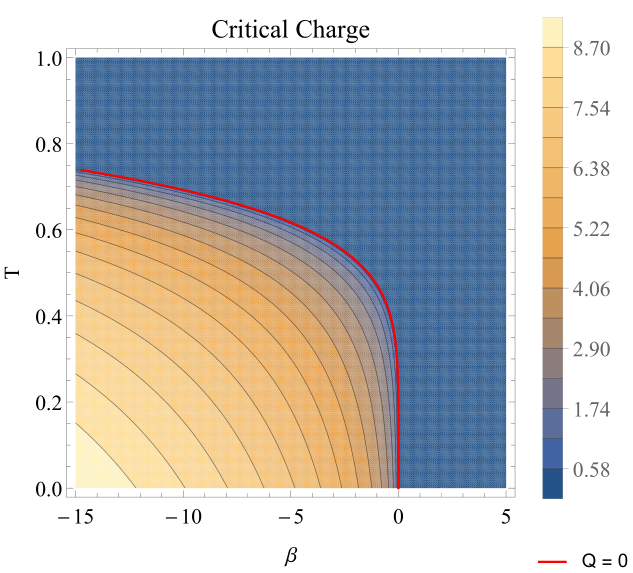}
  \includegraphics[width =0.32\textwidth]{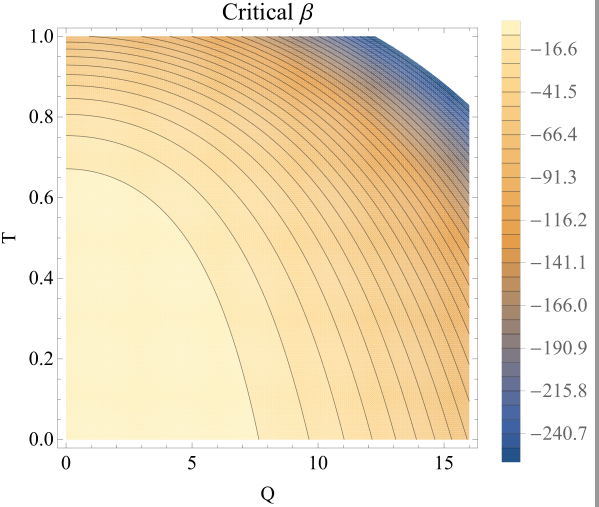}
  \caption{From left to right are the contour plots of critical temperature, critical $Q$ and critical $\beta$.}
  \label{fig:para1}
\end{figure}

We use \texttt{NSolve} in Mathematica to solve the critical values of $(Q,\beta,T)$, then we obtain three contour plots (see Fig. \ref{fig:para1}).
These plots show the critical value of $T,\, Q$, and $\beta$ as functions of the other two parameters, respectively. More specifically, the critical $T$ increases with decreasing $Q$ and $\beta$; the critical $Q$ increases with decreasing $\beta$ and $T$; the critical $\beta$ increases with decreasing $Q$ and $T$. In addition, the red curves in the left and middle plot of Fig. \ref{fig:para1} is the boundary where the critical $T$ or $Q$ vanishes. In general, the critical parameter of each in $(Q,\beta, T)$ decreases along the direction of the other two parameters. Moreover, as mentioned above, when $\beta > 0$, the critical value of both $Q$ and $T$ will drop to $0$. Bounded by these regions, we calculate and discuss the properties of HEE, MI and EWCS in the section \ref{sec:HEE}, \ref{sec:mi} and \ref{sec:eop_phenomena}.

\section{The Holographic entanglement entropy}\label{sec:HEE}

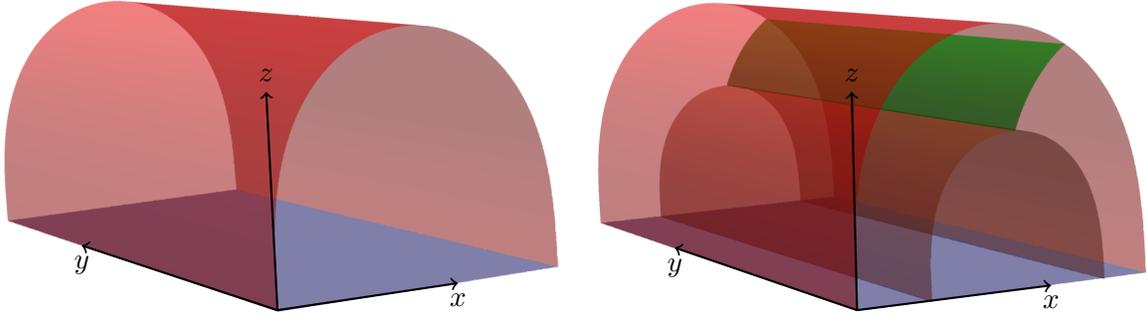
\begin{figure}
  \begin{tikzpicture}[scale=1]
    \node [above right] at (0,0) {\includegraphics[width=7.5cm]{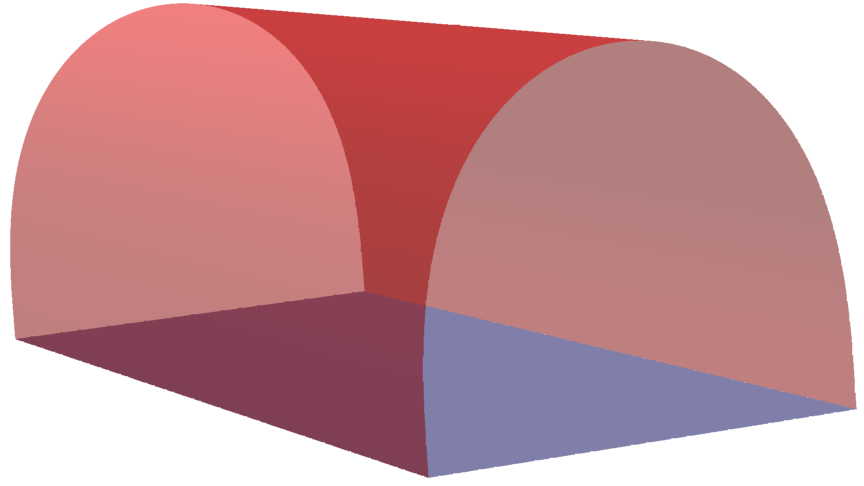}};
    \draw [right,->,thick] (3.85, 0.22) -- (6.25, 0.58) node[below] {$x$};
    \draw [right,->,thick] (3.85, 0.22) -- (1.25, 1.08) node[below] {$y$};
    \draw [right,->,thick] (3.85, 0.22) -- (3.7, 3.125) node[above] {$z$};
  \end{tikzpicture}
  \begin{tikzpicture}[scale=1]
    \node [above right] at (0,0) {\includegraphics[width=7.5cm]{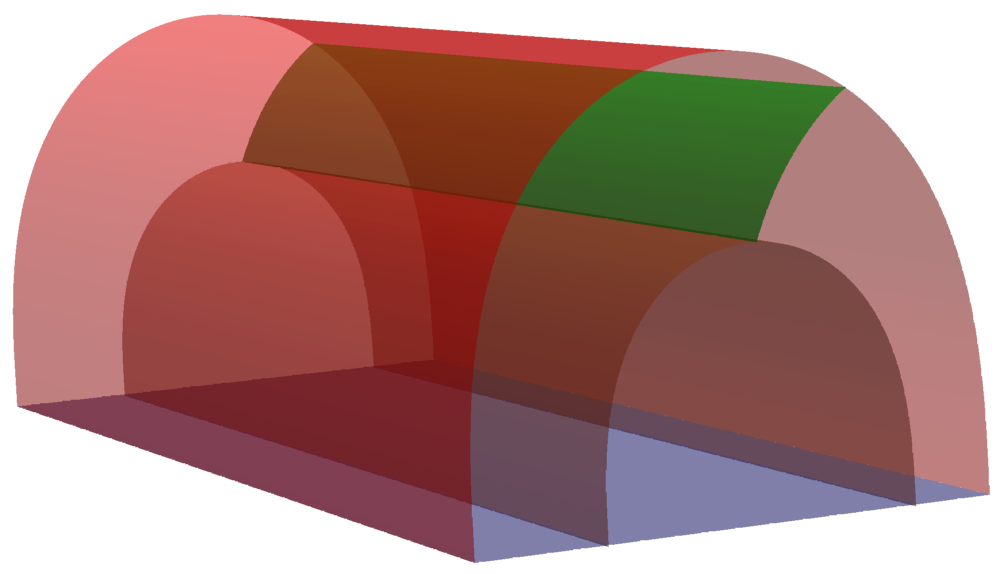}};
    \draw [right,->,thick] (3.67, 0.22) -- (6.25, 0.55) node[below] {$x$};
    \draw [right,->,thick] (3.67, 0.22) -- (1.25, 1.05) node[below] {$y$};
    \draw [right,->,thick] (3.67, 0.22) -- (3.6, 3.125) node[above] {$z$};
  \end{tikzpicture}
  \caption{The left plot: The minimum surface for a given width $w$. The right plot: The minimum cross-section (green surface) of the entanglement wedge.}
  \label{msd1}
\end{figure}

One of the most important features distinguishing quantum systems from classical ones is entanglement, which can be measured by many physical quantities. As the most well-known measure, EE depicts the entanglement between a subsystem and its complement. Given a system composed of disjoint $A$ and $B$, the subsystem $A$ is described by a reduced density matrix $\rho_A = \text{Tr}_{B} \rho_{\text{total}}$. To characterize the entanglement between $A$ and $B$, EE is defined as the von Newmann entropy of the reduced density matrix,
\begin{equation}\label{ee-von}
  S_{A} (|\psi\rangle) = - \text{Tr}\left[ \rho_{A} \log \rho_{A} \right],\quad \rho_{A} = \text{Tr}_{B} \left(|\psi\rangle\langle\psi|\right).
\end{equation}
This definition immediately leads to $S_A = S_B$ for pure states \cite{Chuang:2002book}. Though EE has been widely considered a good entanglement measure for pure states, it is not suitable for describing the entanglement of mixed states. Because even if the degrees of freedom of $A$ and $B$ are not entangled, such as the direct product states, they can still have non-zero EE. Many new entanglement measures have been proposed to characterize the entanglement of mixed states, among which MI is the most commonly used one \cite{vidal:2002,Horodecki:2009review}. In holographic duality theory, the EE was associated with the area of the minimum surface stretching into the bulk of the dual gravity systems (see the left plot of Fig. \ref{msd1}) \cite{Ryu:2006bv}.

In this paper, we focus on the partition of infinite strips along $y$-direction, where solving the minimum surfaces only involves ordinary differential equations. We parametrize the minimum surface with the angle $\theta \equiv \arctan (z/x)$ (see Fig. \ref{fig:cartoon4eop}), which can facilitate the solving of the minimum surface \cite{Liu:2020blk,Liu:2021rks}. The first step of this numerical method is to discretize the angle with Gauss-Lobatoo collocation \cite{Boyd:2001}. Because of the nonlinearity of the equation of motions for the minimum surfaces, we must apply the Newton-Raphson iteration method to find out the minimal surface. Equipped with this numerical method, we study the properties of HEE in the Aether gravity model.

\begin{figure}
  \centering
  \includegraphics[width=0.5\textwidth]{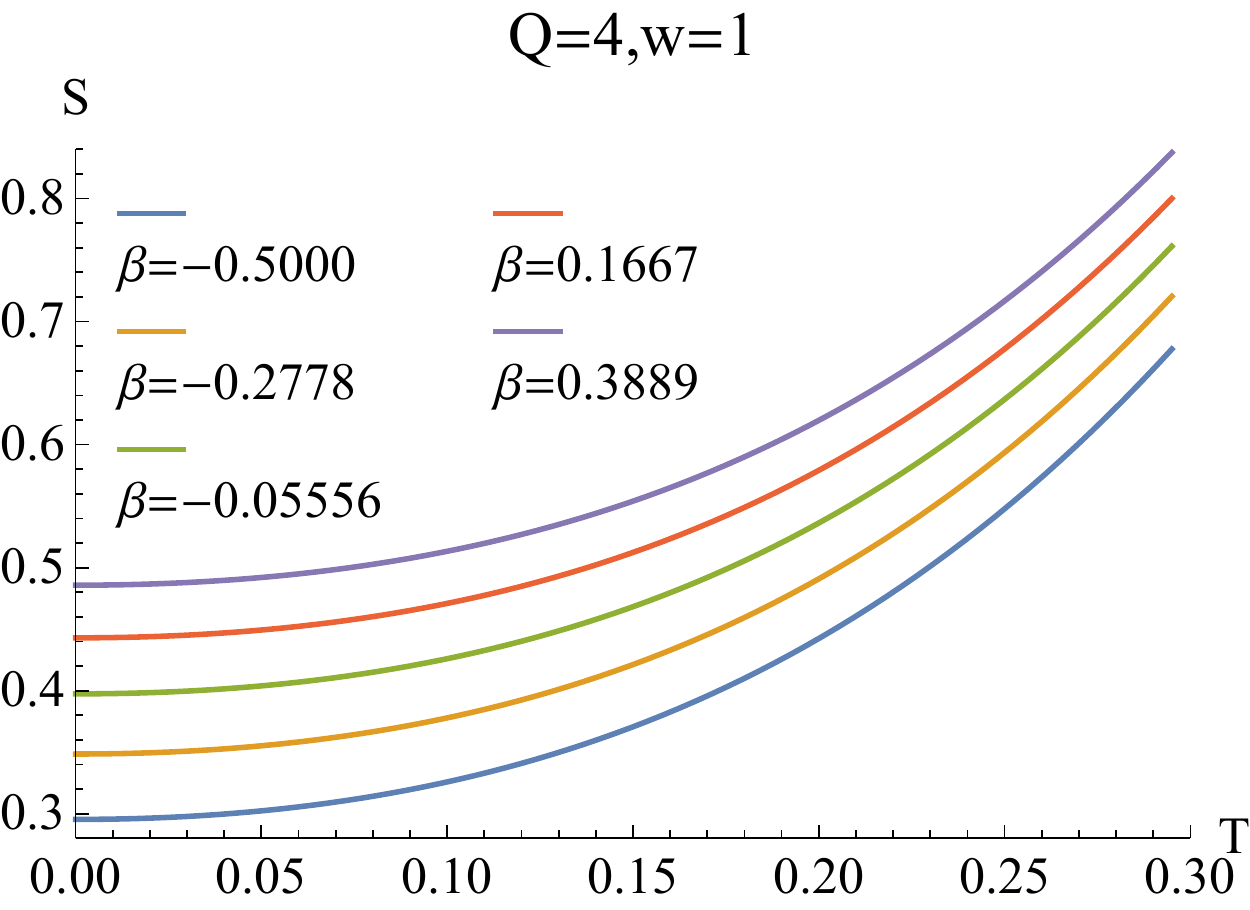}
  \caption{HEE vs $T$ in different value of the $\beta$ when $Q=4$ and width $w=1$.}
  \label{fig:heevsT1}
\end{figure}

First, we show the HEE versus $T$ in Fig. \ref{fig:heevsT1}, from which we can find that HEE increases with $T$, whether the $\beta$ is negative or positive.
The temperature behavior of the HEE depends on the relation between the $r_h$ and the temperature. When the temperature increases, the $r_h$ increases, hence the minimum surface tends to approach the horizon. As a consequence, the HEE increases with the temperature. Physically speaking, this is as expected since the entanglement entropy becomes more and more dominated by the thermal entropy, which monotonically increases with the temperature.

\begin{figure}
  \centering
  \includegraphics[width=0.5\textwidth]{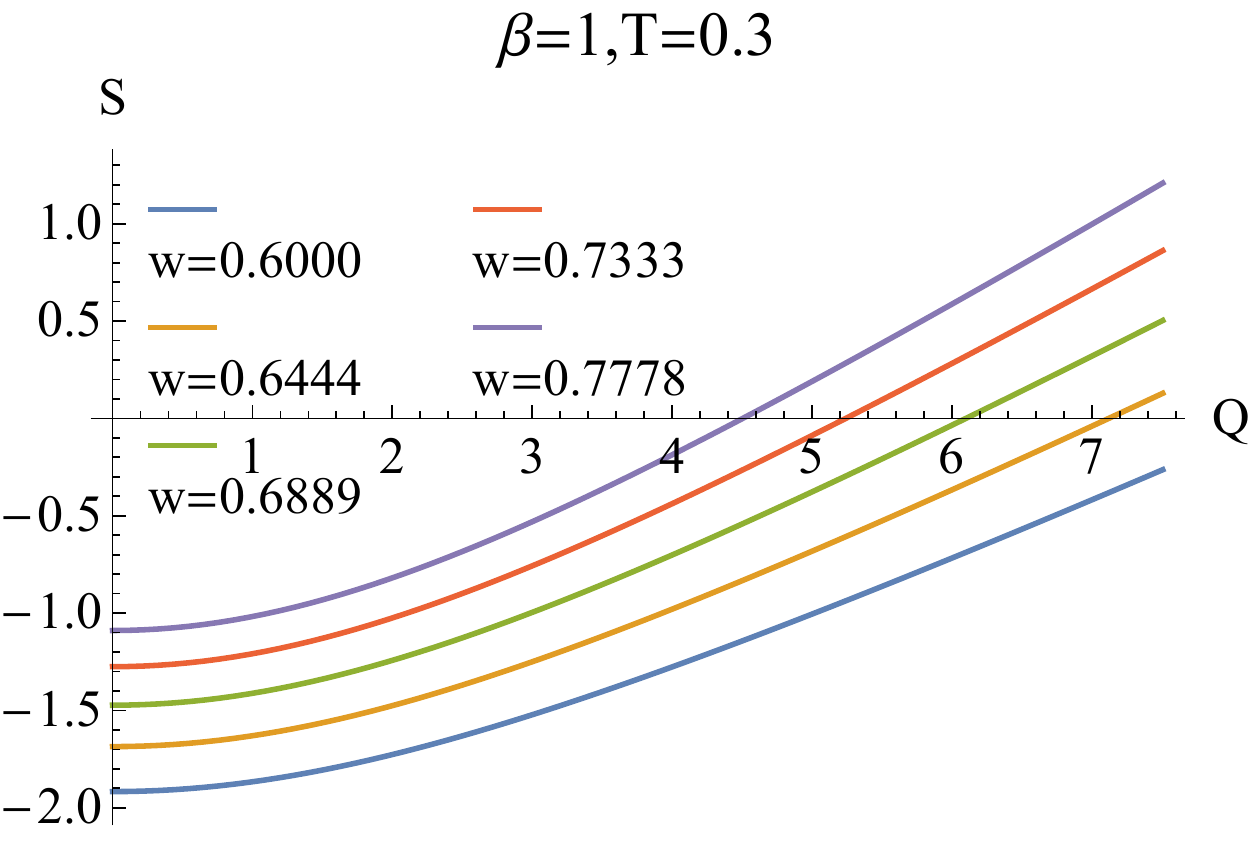}
  \caption{HEE vs $Q$ in different value of width $w$ when $\beta=1$ and $T=0.3$.}
  \label{fig:heevsQ1}
\end{figure}

Next, we show the HEE versus $Q$ in Fig. \ref{fig:heevsQ1}, from which we can find that HEE increases with $Q$ and width $w$.
Similar to the temperature behavior, the HEE behavior with $Q$ can also be attributed to the increase of $r_h$ when increasing $Q$. Similar behaviors have also been obtained in AdS-RN system \cite{Liu:2019qje}.

\begin{figure}
  \centering
  \includegraphics[width=0.5\textwidth]{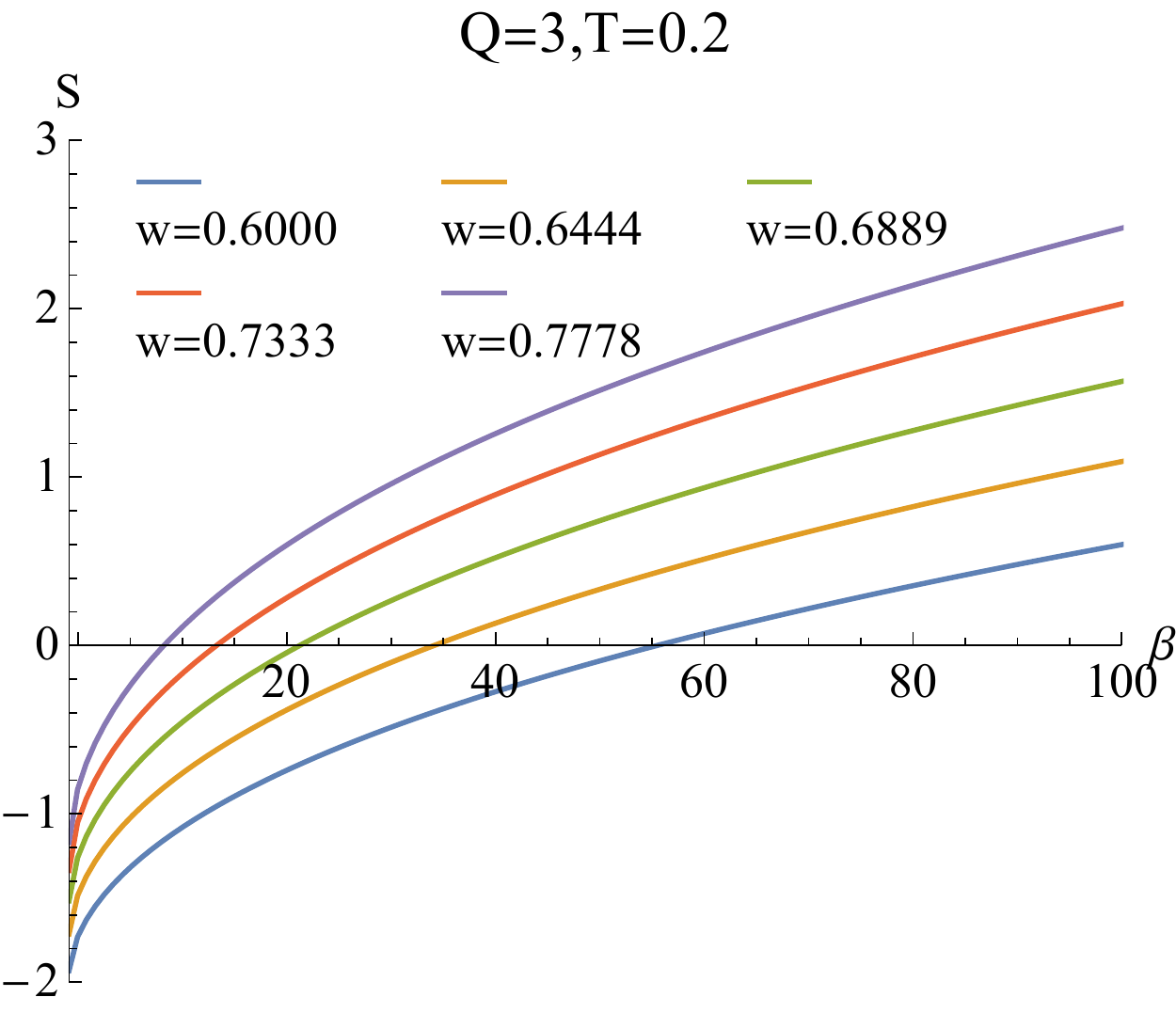}
  \caption{HEE vs $\beta$ in different value of width $w$ when $Q=3$ and $T=0.2$. The HEE increases with the increase of $\beta$.}
  \label{fig:heevsbc1}
\end{figure}

The particularly interesting phenomenon is the HEE along the Lorentz violation parameter $\beta$ (Fig. \ref{fig:heevsbc1}).
The underlying reason for the increasing HEE with $\beta$ is also the same as that of the $T$ and $Q$. The partial derivative of $r_h$ with $\beta$ can be obtained as
\begin{equation}\label{eq:dsdbeta}
  \partial_\beta r_h = \frac{4 r_h}{20 \beta +Q^2 r_h^2+r_h^6},
\end{equation}
which shows that $\partial_\beta r_h >0$ for physically allowed regions. The horizon radius of the black brane becomes larger as the Lorentz violation effect becomes more significant. Since the Lorentz violation effect is induced by the introduction of time-like field $u^a$, it can be clearly seen from Eq. \eqref{eq:Legendreae1} and Eq. \eqref{eq:fandv} that the increase of horizon radius of black brane is caused by the increasing of the action of the matter field $u^a$. Furthermore, it leads to the increase of the entanglement entropy of the dual boundary theory.

Although they have the same monotonicity, it is worth noting that there are obvious differences in the details of HEE behaviors with $Q$, $T$, and $\beta$. First, compared with the increasing behavior of HEE with $\beta$, HEE increases more rapidly with $T$ and $Q$. This phenomenon can be understood from Eq. \eqref{eq:dsdbeta}, where $\partial_\beta r_h$ decreases with the increase of $\beta$. Therefore, with the increase of $\beta$, HEE will increase, meanwhile the slope decreases.

After elaborating on the properties of HEE, we now discuss the properties of MI, the entanglement measure of mixed states, in the Aether gravity.

\section{The Holographic mutual information}\label{sec:mi}

The MI for disjoint $A\cup B$ is defined as
\begin{equation}\label{mi:def}
  I\left(A,B\right) := S\left(A\right) + S\left(B\right) - S\left(A\cup B\right),
\end{equation}
When $\rho_{AB} = \rho_{A} \otimes \rho_{B} $, it can be verified that $ I\left(A,B\right) =0 $. Therefore, MI can recognize that the product states are not entangled. Based on the numerical calculation of HEE in the previous section, we can calculate the MI in aether gravity, as the definition of MI is directly related to HEE.

\begin{figure}
  \centering
  \includegraphics[width=0.5\textwidth]{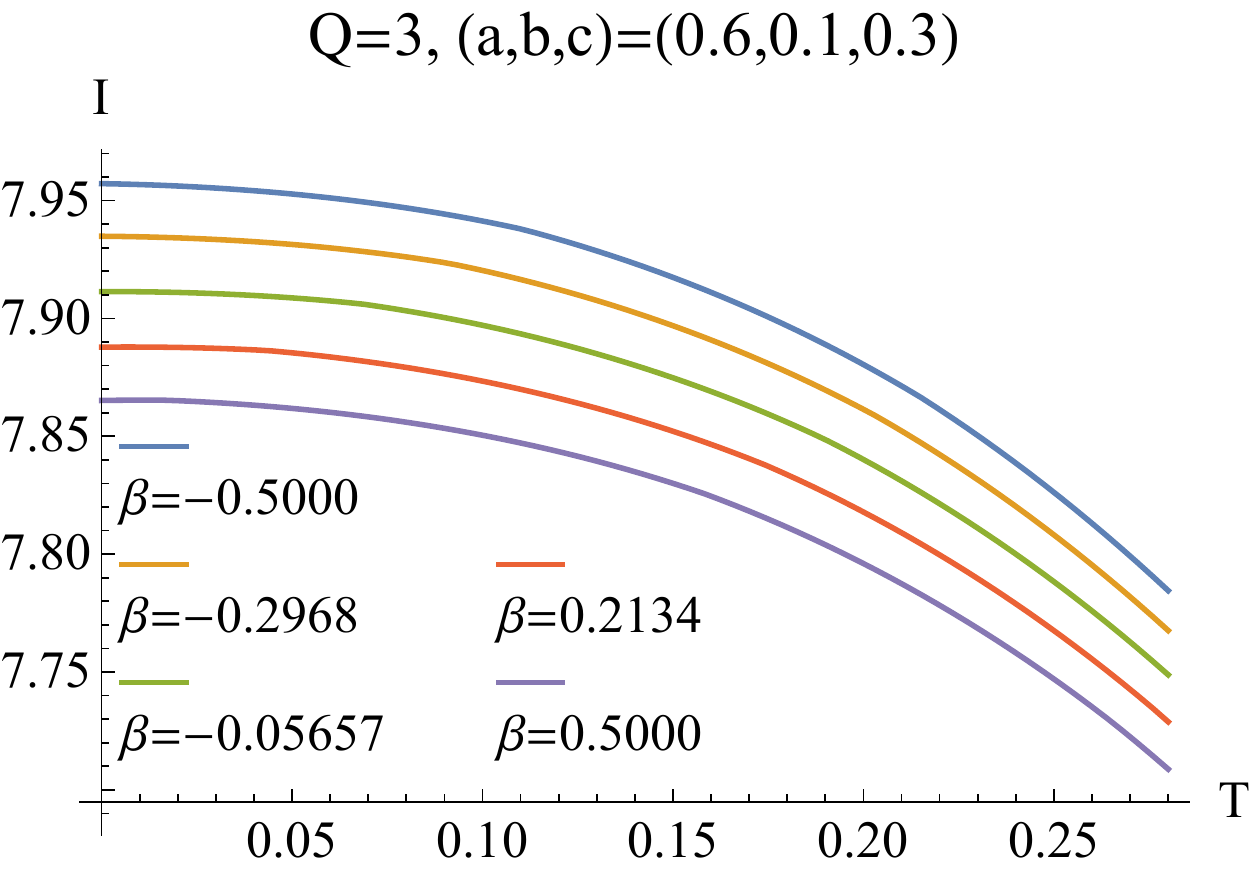}
  \caption{The MI vs $T$ in different value of $T$. It shows the change rule between the MI and two parameters $T$ and $\beta$ that the MI decreases with the increase of $T$ and $\beta$. }
  \label{fig:MIvsT1}
\end{figure}

\begin{figure}
  \centering
  \includegraphics[width=0.6\textwidth]{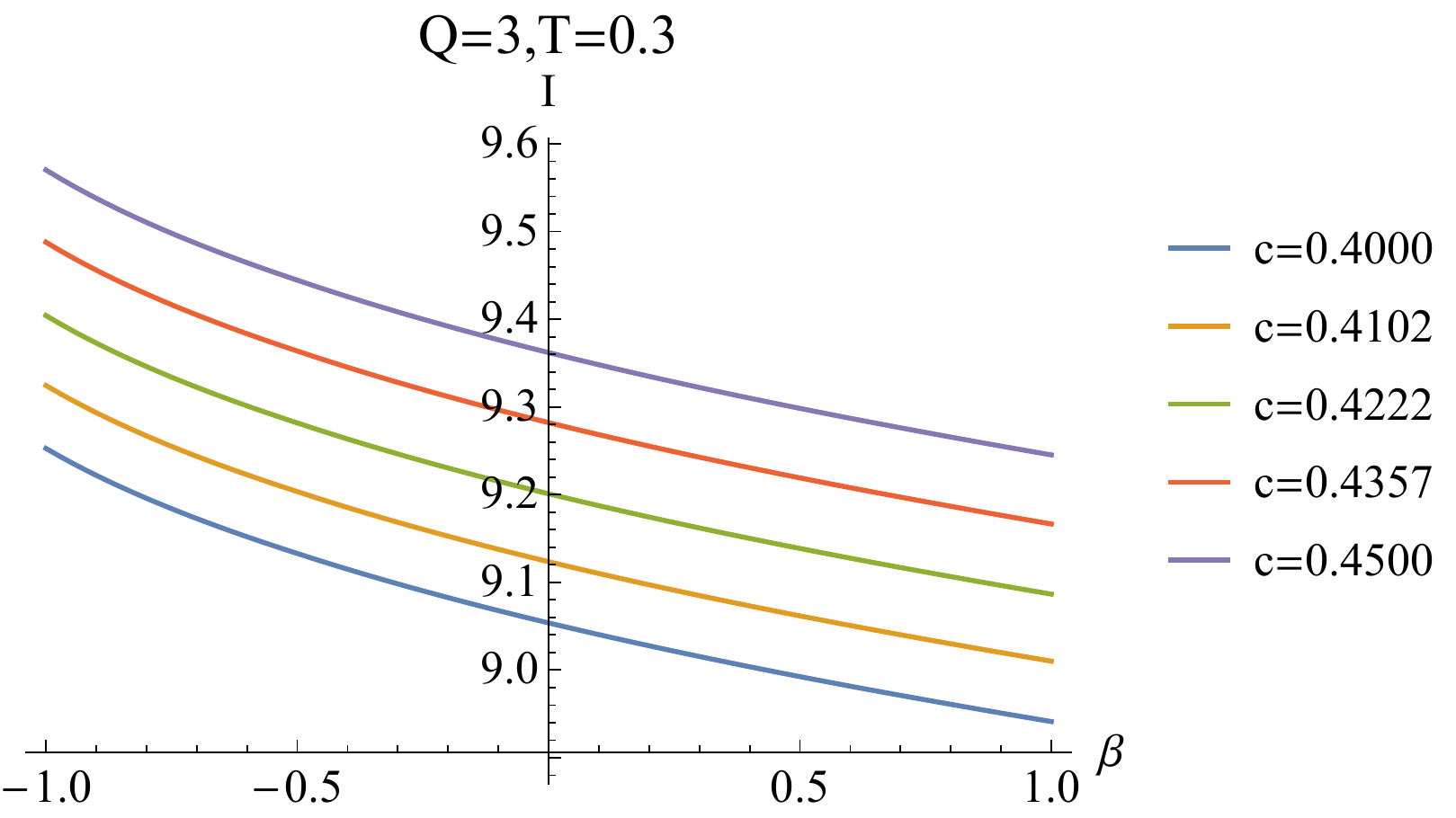}
  \caption{The MI vs $\beta$ in different configurations $(a,b,c)$ by choosing $a=0.6$, $b=0.1$ with different value of $c$.The MI decreses with the increasing $\beta$. However, MI increases with the increasing of c.This means that MI increases when the configuration tends to be symmetrical.}
  \label{fig:MIvsbc1}
\end{figure}

\begin{figure}
  \centering
  \includegraphics[width=0.8\textwidth]{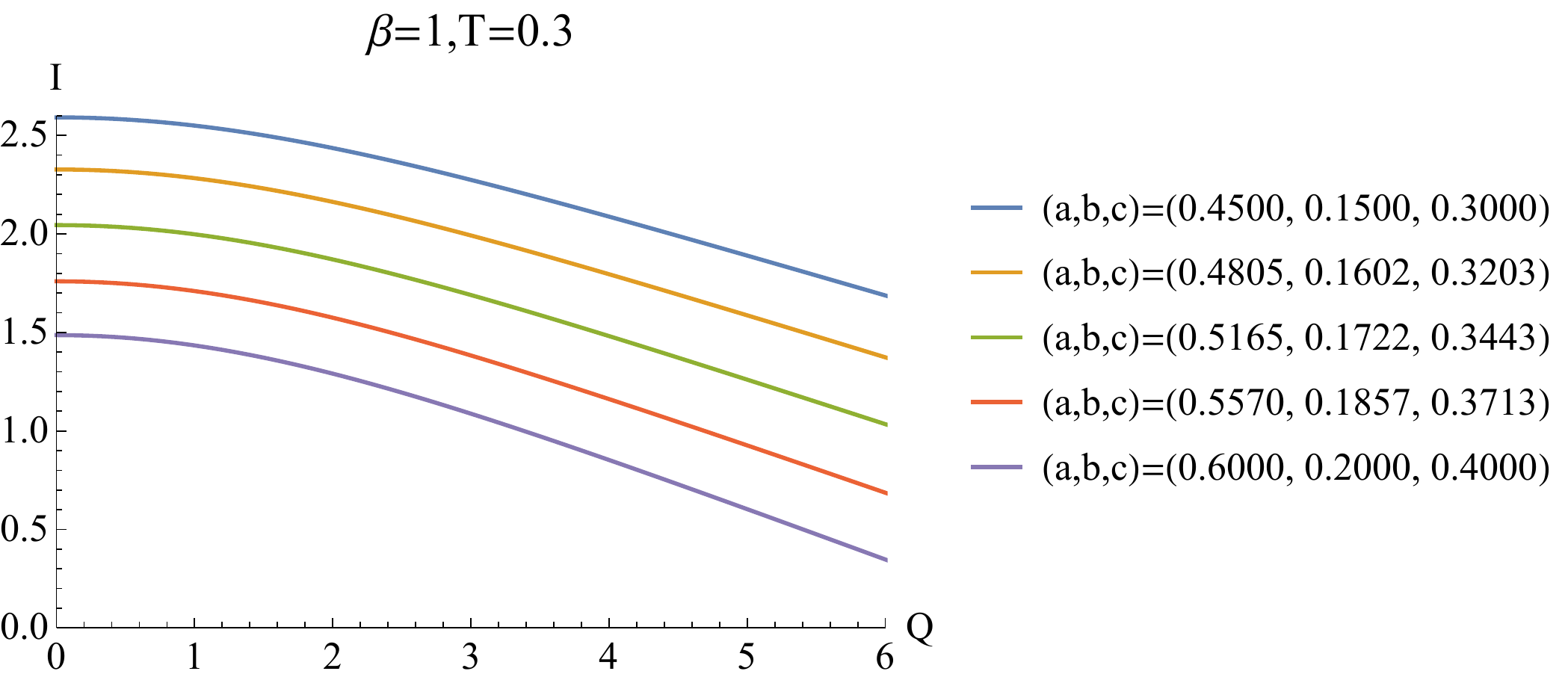}
  \caption{The MI vs $Q$ in different configurations $(a,b,c)$ taking the same proportion between $a$ ,$b$ and $c$. The MI decreases with the increasing charge $Q$ and the larger $(a,b,c)$.}
  \label{fig:MIvsQ1}
\end{figure}

We demonstrate the behavior of MI in the Aether gravity as follows.
In general, given a configuration $(a,b,c)$, the behavior of MI is opposite to that of HEE - MI decreases with the increase of $T$, $Q$ and $\beta$. First, we can see from Fig. \ref{fig:MIvsT1}, Fig. \ref{fig:MIvsbc1} and Fig. \ref{fig:MIvsQ1} that MI always decreases with the increase of $T,\,\beta,\,Q$. The reason responsible for the opposite monotonic behavior of the MI to that of the HEE directly results from the HEE. According to the definition \eqref{mi:def}, a non-trivial MI will be $I=S_A + S_C - S_B - S_{A\cup B\cup C}$, the last term $S_{A\cup B\cup C}$ is more affected by the deformation caused by the deviation from the AdS, and therefore the MI shows the opposite behavior.

Moreover, we can find from Fig. \ref{fig:MIvsbc1} that MI increases with the increase of $c$ when fixing $a$ and $b$. The reason behind this phenomenon is that increasing the size of the subregion can increase the degrees of freedom that can entangle with other degrees of freedom. In addition, we discuss the behavior of MI when varying the sizes of $a$, $b$, and $c$ in equal proportions (see Fig. \ref{fig:MIvsQ1}), where different curve corresponds to different value of $(a,b,c)$. The MI decreases when uniformly increasing the values of $(a,b,c)$. This shows that the decrease of entanglement caused by the increase of the separation plays a major role when increasing the subregions and the separation in the same proportion. In fact, this is a reasonable result. Because the entanglement between the degrees of freedom usually decays more rapidly with the increase of the separation, when the separation increases to a certain value, the subregions will be disentangled.

The above phenomena show that MI is directly determined by HEE, thus MI may not be a good measure of mixed state entanglement. Therefore we need to resort to other mixed state entanglement measures. In the next section, we will study a new entanglement measure, the EWCS, in the Aether gravity.

\section{The Entanglement wedge minimum cross-section}\label{sec:eop_phenomena}

The entanglement wedge minimum cross-section (EWCS) has been associated with several different mixed state entanglement measures, such as entanglement of purification, the reflected entropy, odd entropy, and so on. It involves the purification process of mixed states. When the entanglement wedge exists (i.e., where MI is non-trivial), the mixed entanglement measures $ E_{W}\left(\rho_{AB}\right) $ is proportional to the area of the minimum cross-section $ \Sigma_{AB} $ in it \cite{Takayanagi:2017knl},
\begin{equation}\label{heop:def}
  E_{W}\left(\rho_{AB}\right) = \min_{\Sigma_{AB}} \left( \frac{\text{Area} \left(\Sigma_{AB}\right)}{4G_{N}}\right).
\end{equation}
EWCS vanishes when the entanglement wedge becomes disconnected, i.e., when MI vanishes.

Though the explicit definition is given, the following three reasons explain the difficulty of solving the EWCS. First, the equations of motion of minimum surfaces are highly nonlinear, which are usually hard to solve. Second, the minimum cross-section is hard to find because each cross-section itself is already a local minimum, which means that minimizing the cross-section is a second-order minimization. Last but not least, the coordinate singularity at the horizon will easily sabotage the numerical precision, adding to the difficulty of our numerics.

\begin{figure}
  \centering
  \includegraphics[width = \textwidth]{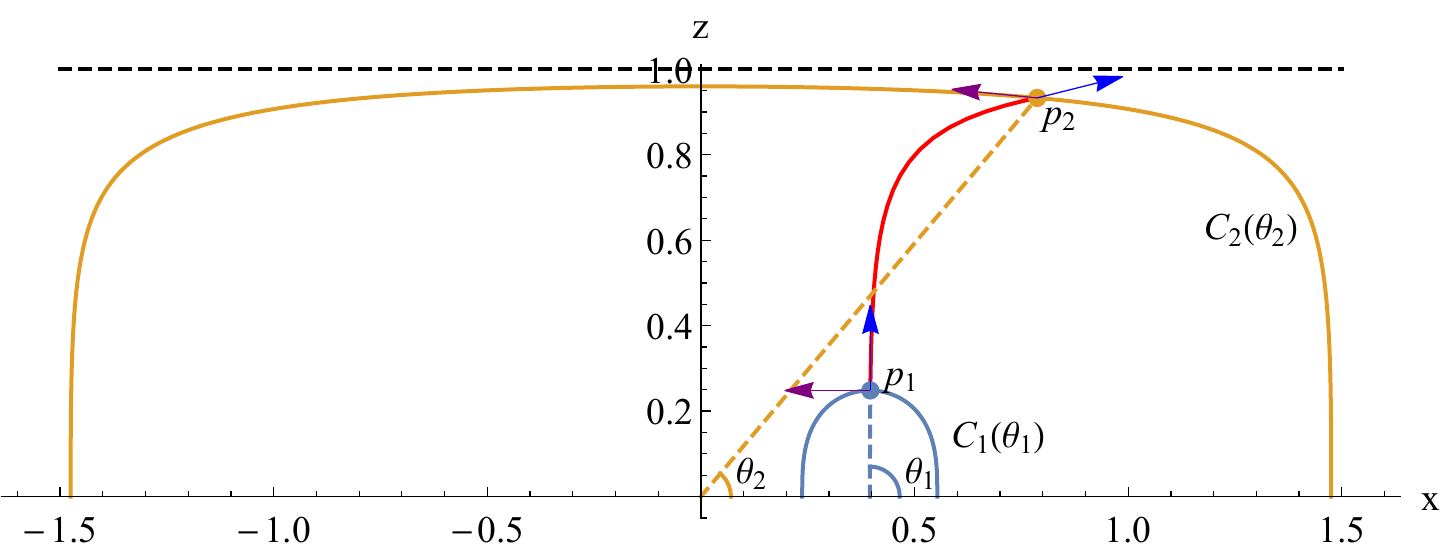}
  \caption{
    Schematic diagram of solving EWCS. The inner blue curve $C_1(\theta_1)$ and the outer orange curve $C_2(\theta_2)$ in the figure are the corresponding minimum surfaces with widths of $b$ and $a + b + c$ respectively. The red curve is the minimum surface connecting $p_1$ on $C_1$ and $p_2$ on $C_2$. There are blue and purple arrows at $p_1$ and $p_2$, representing the tangent vector along the red curve and the tangent vector along $C_1$ or $C_2$. The horizontal black dashed line represents the horizon of the black brane.
  }
  \label{fig:cartoon4eop}
\end{figure}

An efficient algorithm for solving EWCS has been developed by using the boundary condition that the minimum cross-section must be locally orthogonal to the entanglement wedge \cite{Liu:2020blk}. Fig. \ref{fig:cartoon4eop} shows a schematic diagram of the main ideas of the algorithm for solving EWCS.
We focus on the EWCS of parallel infinite strips in a homogeneous background,
\begin{equation}\label{genbg}
  ds^{2} = {g_{tt}} dt^2 + g_{zz}dz^2 + g_{xx}dx^2 + g_{yy}dy^2,
\end{equation}
where the asymptotic AdS boundary resides at $z = 0$.
The homogeneity implies that $z$ is the only variable of the metric components $g_{\mu\nu}$. For a biparty subsystem with minimum surfaces $C_1(\theta_1),\,C_2(\theta_2)$, we work out the minimum surface $C_{p_1,p_2}$ connecting $p_1 \in C_1$ and $p_2\in C_2$. By parametrizing $C_{p_1,p_2}$ with $z$, the area of $C_{p_1,p_2}$ reads,
\begin{equation}\label{eq:zpara}
  A = \int_{C_{p_1,p_2}} \sqrt{ g_{xx} g_{yy} x'(z)^2 + g_{zz} g_{yy} } dz.
\end{equation}
The equation of motion from minimizing $A$ reads,
\begin{equation}\label{eq:zparaeom}
  x'(z)^3 \left(\frac{g_{ xx } g_{ yy }'}{2 g_{ yy } g_{ zz }}+\frac{g_{ xx }'}{2 g_{ zz }}\right)+x'(z) \left(\frac{g_{ xx }'}{g_{ xx }}+\frac{g_{ yy }'}{2 g_{ yy }}-\frac{g_{ zz }'}{2 g_{ zz }}\right)+x''(z) =0,
\end{equation}
with boundary conditions,
\begin{equation}\label{eq:zparabcs}
  x(z(\theta_i)) = x(\theta_i),\quad i=1,2.
\end{equation}

The local orthogonal relation between the minimum cross-section and the entanglement wedge leads to,
\begin{equation}\label{eq:perpend}
  \left\langle \frac{\partial}{\partial z},\frac{\partial}{\partial \theta_1} \right\rangle_{p_1} = 0,\quad \left\langle \frac{\partial}{\partial z},\frac{\partial}{\partial \theta_2} \right\rangle_{p_2} = 0,
\end{equation}
where $\langle\cdot,\cdot\rangle$ represents the vector product under the metric $g_{ab}$. In order to control the numerical precision, we adopt the normalized local orthogonal relation,
\begin{equation}\label{eq:perpend2}
  Q_1(\theta_1,\theta_2) \equiv \left.\frac{ \left\langle \frac{\partial}{\partial z},\frac{\partial}{\partial \theta_1} \right\rangle }{\sqrt{ \left\langle \frac{\partial}{\partial z},\frac{\partial}{\partial z} \right\rangle \left\langle \frac{\partial}{\partial \theta_1},\frac{\partial}{\partial \theta_1} \right\rangle }}\right|_{p_1} = 0, \quad
  Q_2(\theta_1,\theta_2)  \equiv \left.\frac{ \left\langle \frac{\partial}{\partial z},\frac{\partial}{\partial \theta_1} \right\rangle }{\sqrt{ \left\langle \frac{\partial}{\partial z},\frac{\partial}{\partial z} \right\rangle \left\langle \frac{\partial}{\partial \theta_2},\frac{\partial}{\partial \theta_2} \right\rangle }}\right|_{p_2} = 0.
\end{equation}
Now, the EWCS can be given when we find the minimum surface on which $(\theta_1,\theta_2)$ of the endpoints satisfy \eqref{eq:perpend2}. To this end, we variate the endpoints satisfying the local perpendicular conditions by the Newton-Raphson method. Next, we study the relationship between the Aether gravity and the EWCS based on the above techniques.

\begin{figure}
  \centering
  \includegraphics[width=0.7\textwidth]{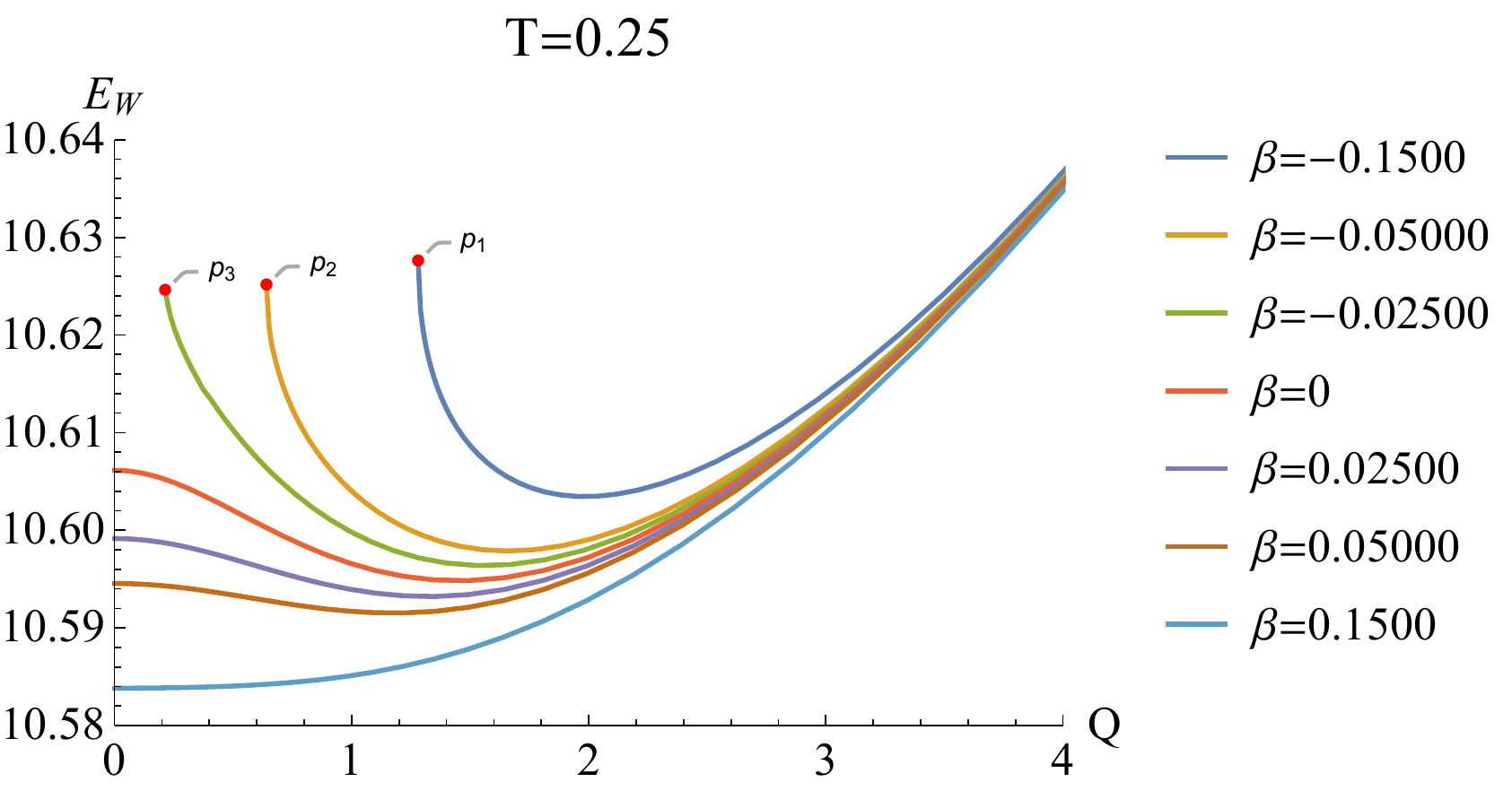}
  \caption{The EWCS vs $Q$ in different value of $\beta$ and the configuration $(a,b,c)$ is $(0.6,0.1,0.3)$. The EWCS behave non-monotonicity with the increase of charge $Q$, which also will vanish if $\beta$ takes a rather larger value. Three red points represent the critical charge $Q$.}
  \label{fig:eopvsQ1}
\end{figure}

\begin{figure}
  \centering
  \includegraphics[width=0.6\textwidth]{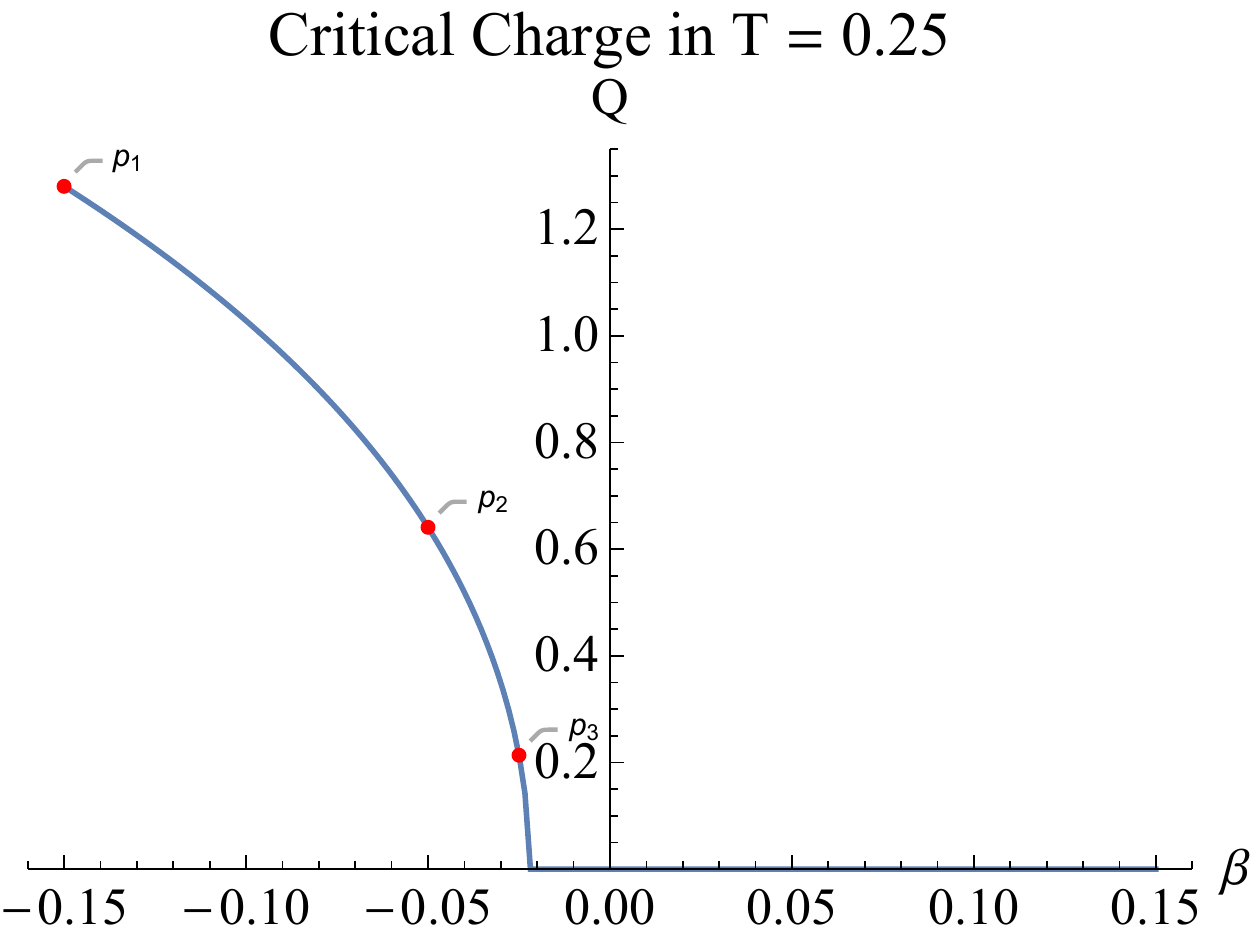}
  \caption{Critical charge $Q$ in case of $T = 0.25$}
  \label{fig:paraQbc1}
\end{figure}

The EWCS behaves distinctly from the HEE or MI in Aether gravity. We fix the configuration $(a,b,c)$ to $(0.6,0.1,0.3)$ since the main phenomenon is independent of this. First, the EWCS shows non-monotonicity with the increase of $Q$ (see Fig. \ref{fig:eopvsQ1}).
EWCS decreases with the increase of $Q$ at the beginning and then starts to increase. However, when $\beta$ increases to $0.15$, we can see that EWCS increases monotonically with the increase of $Q$. The non-monotonicity of it will vanish if the $\beta$ takes a larger value. In addition, three red points $p_1$, $p_2$ and $p_3$ in Fig. \ref{fig:paraQbc1} show the critical charge $Q$ of the cases $\beta=-0.15,-0.05,-0.025$ in Fig. \ref{fig:eopvsQ1} subject to the allowed parameter region shown in Fig. \ref{fig:para1}.

\begin{figure}
  \centering
  \includegraphics[width=0.6\textwidth]{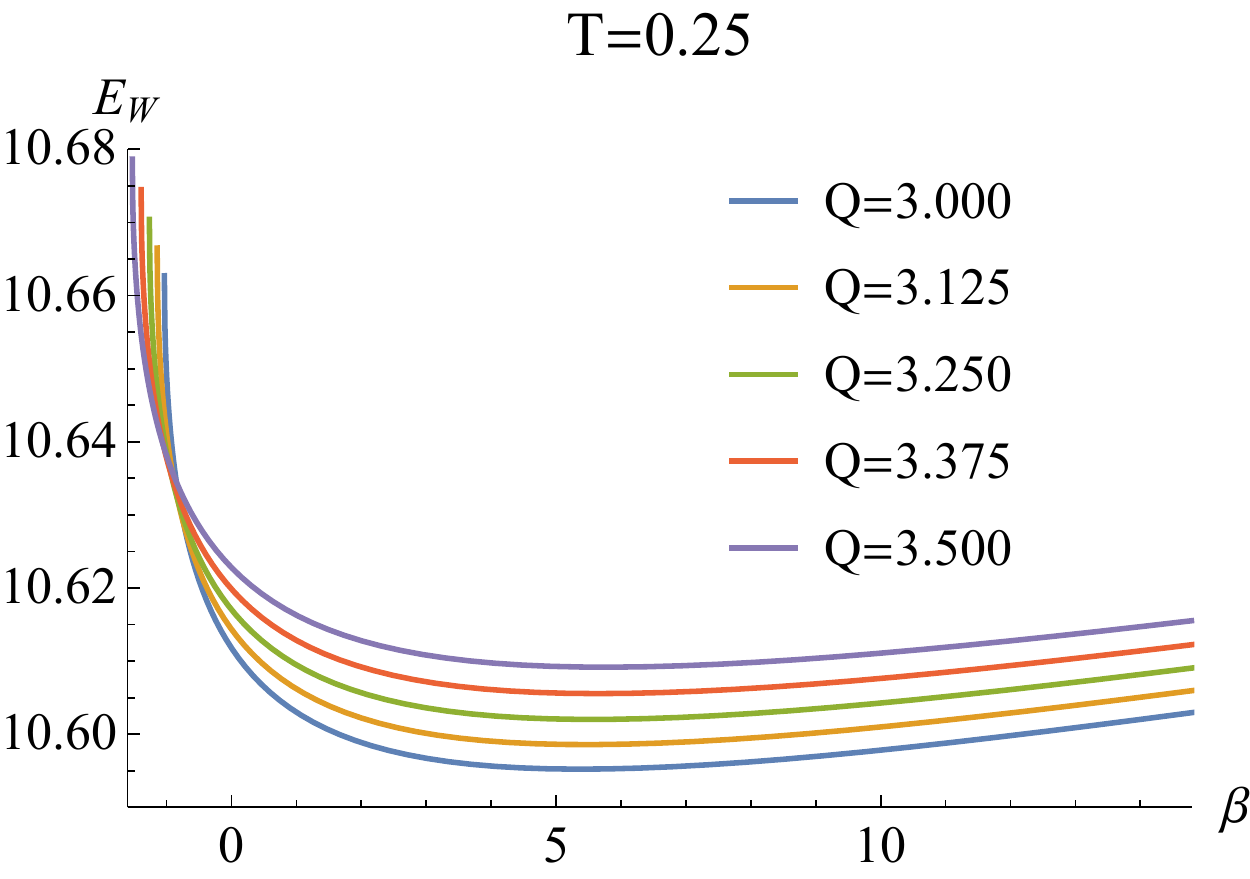}
  \caption{The EWCS vs $\beta$ in different value of $Q$ and the configuration $(a,b,c)$ is $(0.6,0.1,0.3)$.The EWCS shows the non-monotonicity with the increase of $\beta$ at every $Q$. We can see that the EWCS first decreases with $Q$ and $\beta$, and then turns into an increase.}
  \label{fig:eopvsbc1}
\end{figure}

\begin{figure}
  \centering
  \includegraphics[width=0.6\textwidth]{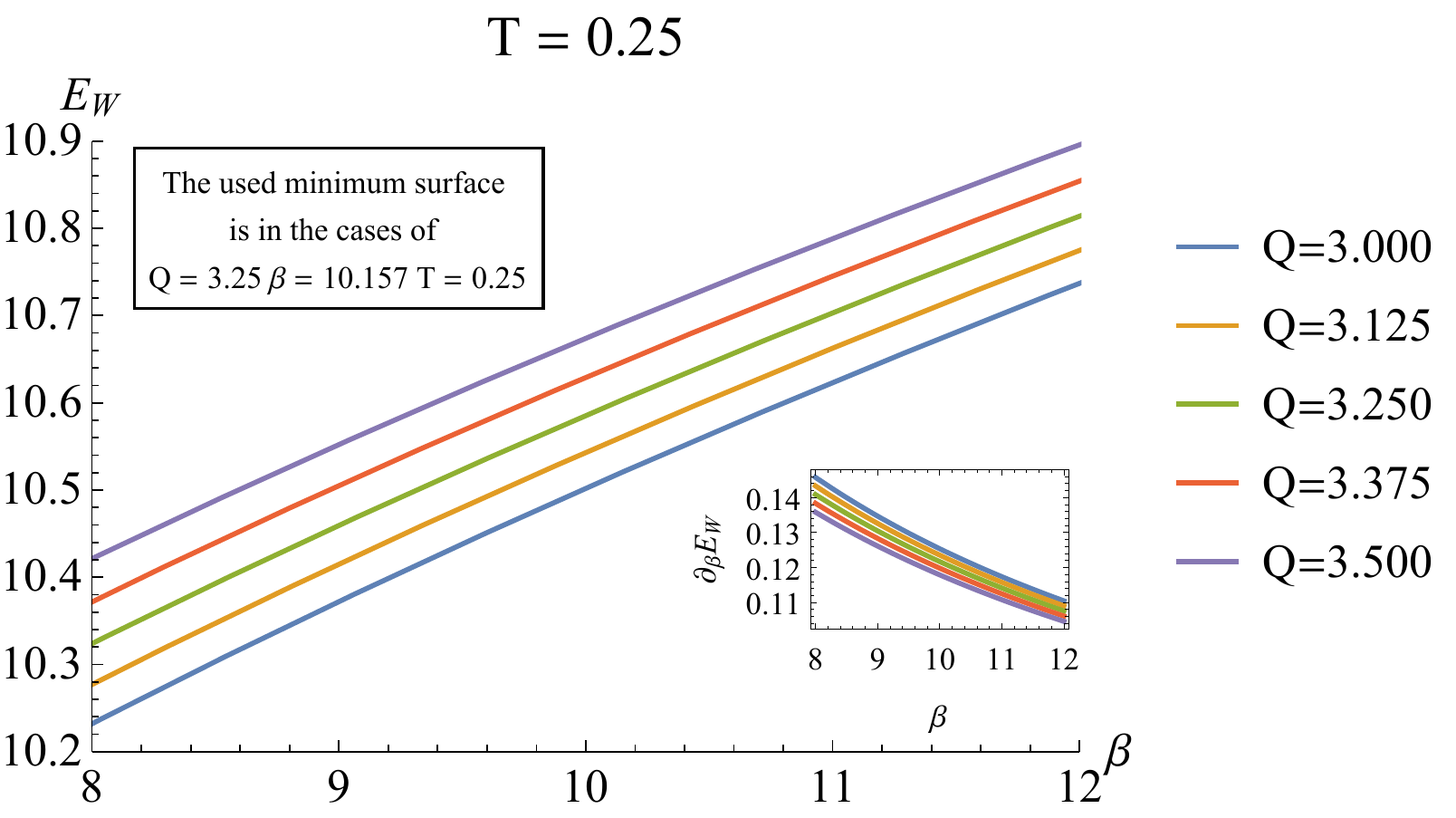}
  \caption{
    The EWCS vs $\beta$ in different values of $Q$ and the configuration $(a,b,c)$ is $(0.6,0.1,0.3)$ when fixing the minimum surface and varying the background geometry.
  }
  \label{fig:cal12}
\end{figure}

For EWCS behavior along $\beta$-direction, the EWCS first decreases and then increases with the increasing $\beta$ (see Fig. \ref{fig:eopvsbc1}).
Also, the intersections shown in Fig. \ref{fig:eopvsbc1} again reflects the non-monotonicity of EWCS with $Q$. When $\beta$ is small, EWCS decreases rapidly with the growth of $\beta$, which can also be understood through the relationship between $r_h$ and $\beta$. It is noted that $r_h$ increases rapidly with the increase of $\beta$ when $\beta$ is small. When $r_h$ increases, the minimum surface will be close to the horizon of the black hole and moves away from the AdS boundary. Because the area of EWCS is mainly contributed by AdS boundary, thus when $\beta$ is small, EWCS will decrease rapidly with the increase of $\beta$. With the increase of $\beta$, $r_h'(\beta)$ becomes smaller, so the growth of EWCS slows down gradually. Therefore, EWCS will present a relatively flat area. When $\beta$ is large, the increasing behavior of EWCS with the increase of $\beta$ is no longer controlled by the deviation from AdS. Instead, we must consider the specific contribution of bulk geometry. Specifically, the variation of the EWCS comes from the variation of the background metric and the minimum surface,
\begin{equation}\label{eq:varia1}
  \begin{aligned}
    \delta E_W & = \delta \int_{ C_{p_1,p_2} } \sqrt{g_{yy} \left( g_{xx} dx^2 + g_{rr} dr^2 \right) }                                                                          \\
               & = \int_{ C_{p_1,p_2} } \frac{\delta E_W}{\delta g_{\mu\nu}} \delta g_{\mu\nu} + \int_{ C_{p_1,p_2} } \frac{\delta E_W}{\delta C_{p_1,p_2}} \delta C_{p_1,p_2}.
  \end{aligned}
\end{equation}
The $\frac{\delta E_W}{\delta C_{p_1,p_2}}$ of \eqref{eq:varia1} is the equation of motion, which must vanish. The first term on the right-hand side of \eqref{eq:varia1} denotes the contribution to the EWCS from the variation of the background metric, which results from the variation of the $F_a(r)$.
\begin{equation}\label{eq:dewdg}
  \begin{aligned}
    \partial_\beta E_W & = \int_{ C_{p_1,p_2} } \partial_\beta g_{rr} \sqrt{\frac{g_{yy}}{ g_{xx}dx^2 + g_{rr}dr^2 }}  dr^2                                   \\
                       & = \int_{ C_{p_1,p_2} } \left( -\frac{\partial_\beta F_a(r)}{F_a^2(r)} \right) \sqrt{\frac{g_{yy}}{ g_{xx}dx^2 + g_{rr}dr^2 }}  dr^2.
  \end{aligned}
\end{equation}
After replacing $M_a$ and $Q$ with horizon boundary conditions and Hawking temperature, we obtain that,
\begin{equation}\label{eq:fa2}
  F_a(r) = -\frac{-r^4 r_h+4 \pi  T r_h^4-3 r_h^5+r r_h^4+Q^2 r}{r^2 r_h}-\frac{4 \beta  \left(3 r^2 r_h-r_h^3+r^3\right)}{r^4 r_h^3}.
\end{equation}
Together with the expression $\partial_\beta r_h$ \eqref{eq:dsdbeta}, we will find that,
\begin{equation}\label{eq:dfdbeta}
  \partial_\beta F_a(r) = \frac{4}{5 r^4 r_h^3} \left(\frac{2 r^2 r_h^2 \left(3 r_h^4 \left(-11 r_h+2 r+10 \pi  T\right)-Q \left(3 r_h+4 r\right)\right)}{20 \beta +Q r_h^2+r_h^6}-9 r^2 r_h+5 r_h^3-2 r^3\right).
\end{equation}
When $\beta$ is relatively large,
\begin{equation}\label{eq:dfdbeta}
  \partial_\beta F_a(r) \simeq  -\frac{4 \left(9 r^2 r_h-5 r_h^3+2 r^3\right)}{5 r^4 r_h^3} <0,
\end{equation}
and hence the second term in \eqref{eq:fa2} is negative and $\partial_\beta E_W >0$ in \eqref{eq:dewdg}. This arguments are also supported by numerics shown in Fig. \ref{fig:cal12}, where varying only the metric will render a increasing behavior of the EWCS with $\beta$.

From the dual picture, when the Lorentz symmetry violation effect is weak, the mixed state entanglement decreases rapidly with the enhancement of Lorentz symmetry violation; when the Lorentz symmetry violation effect is strong, the mixed state entanglement of the system increases slowly when further enhancing the Lorentz symmetry violation.

\begin{figure}
  \centering
  \includegraphics[width=0.45\textwidth]{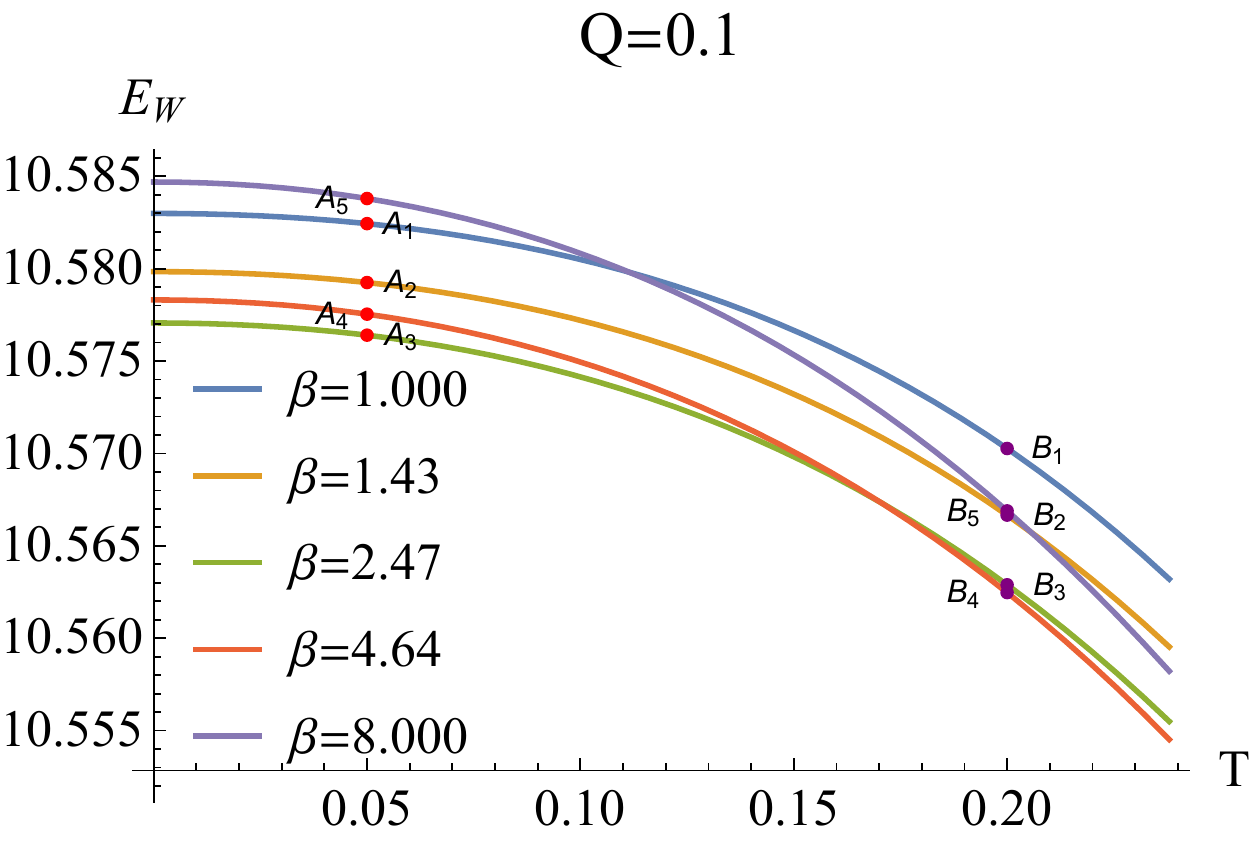}
  \includegraphics[width=0.45\textwidth]{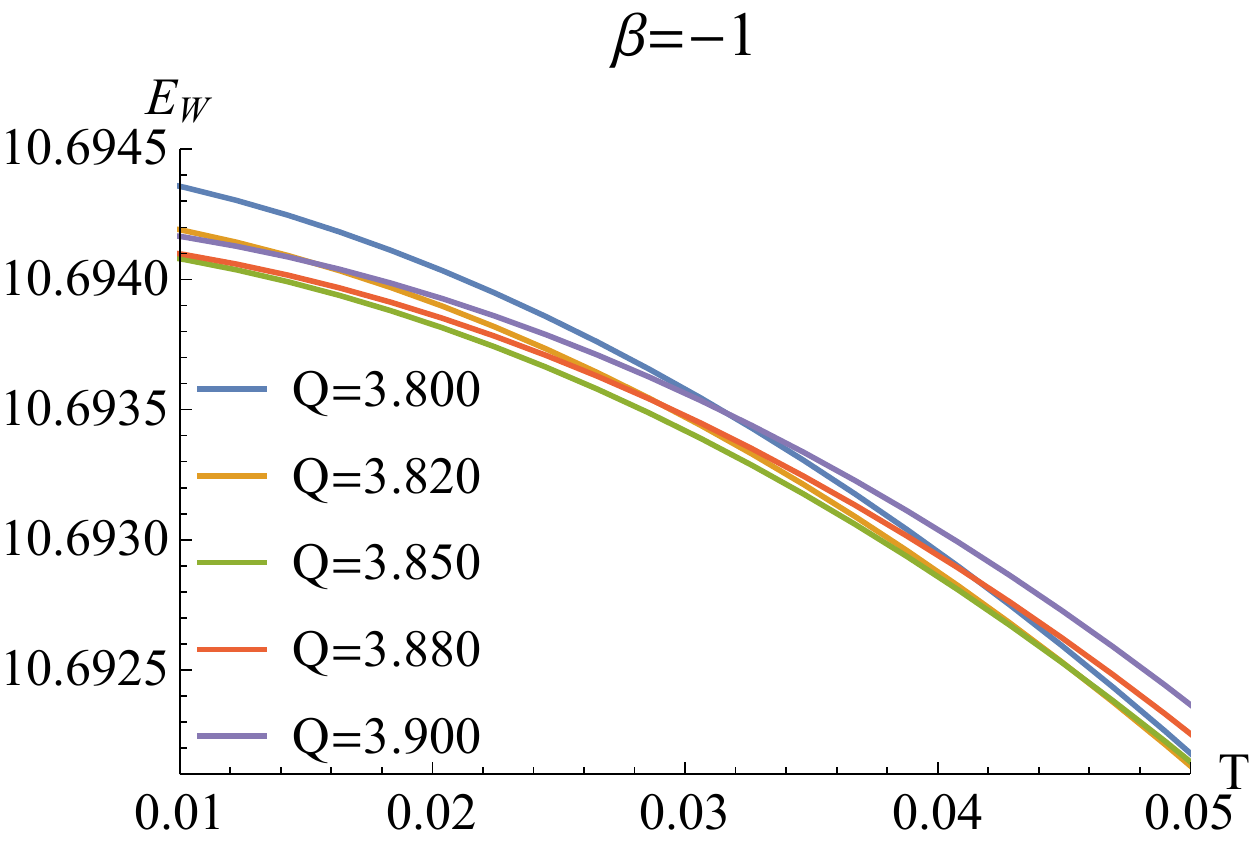}
  \caption{The former plot is EWCS vs $T$ in different values of $\beta$. The EWCS decrease with the increasing temperature. There are several red points and purple points that show the non-monotonicity of the EWCS vs $\beta$. The latter plot is EWCS vs $T$ in different values of $Q$. The configuration $(a,b,c)$ of them are both $(0.6,0.1,0.3)$}
  \label{fig:eopvsT1}
\end{figure}

\begin{figure}
  \centering
  \includegraphics[width=0.5\textwidth]{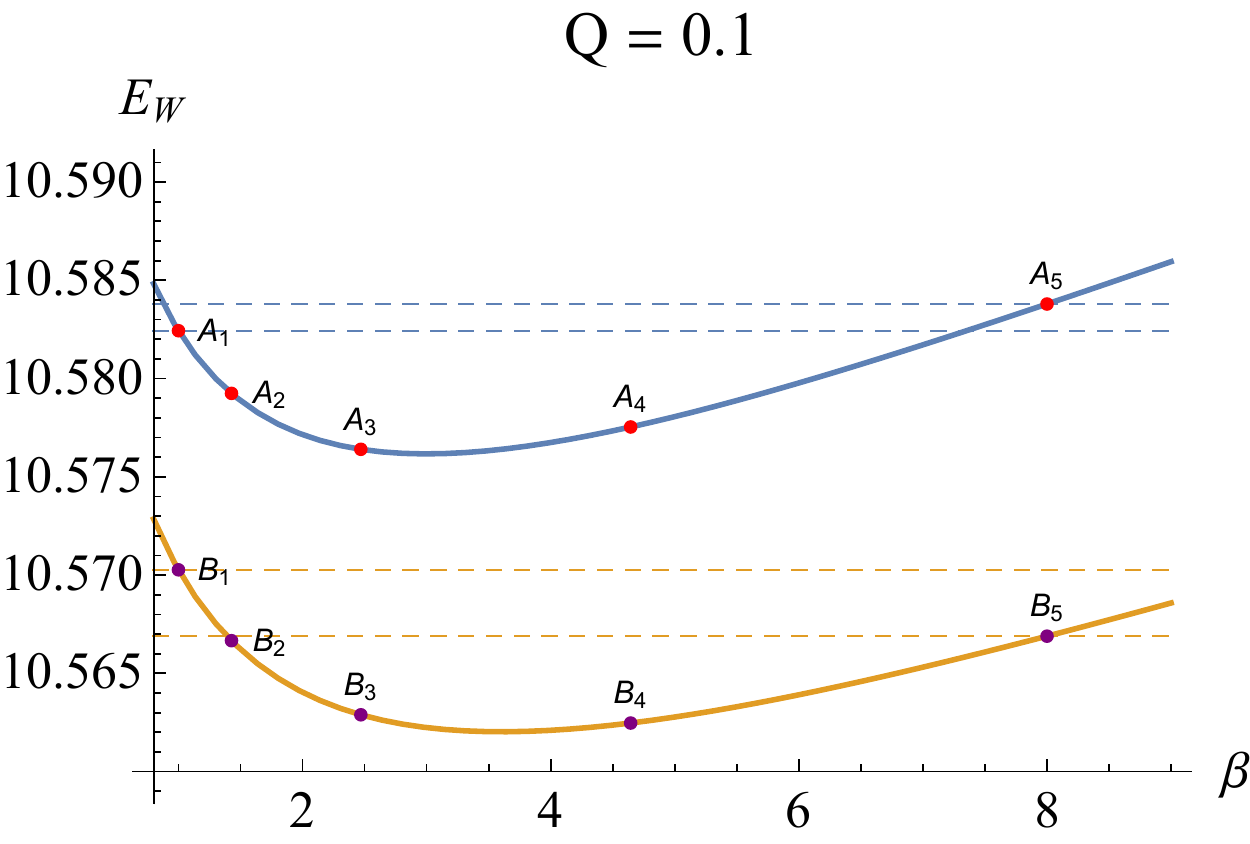}
  \caption{The EWCS vs $\beta$ in different value of $T$ in cases of $T = 0.05$ and $T = 0.2$.}
  \label{fig:eopvsT1C}
\end{figure}

In Fig. \ref{fig:eopvsT1}, EWCS decreases with the increase of temperature. This is in line with physical expectations. Usually, heating up a quantum system will destroy the quantum entanglement between the two subregions. Therefore, EWCS behavior with the temperature here will reflect this.
Also, in the left plot of Fig. \ref{fig:eopvsT1} the curves with different values of $\beta$ cross each other.
Specifically, we can see the red points and the purple points on curves, which correspond to the points in Fig. \ref{fig:eopvsT1C}, from which it is easy to see that EWCS behaves non-monotonically with $\beta$ at a fixed temperature.
The point $A_1$ is lower than the point $A_5$ in case of $T = 0.5$, however when changing the temperature to $T = 0.2$, the point $B_1$ become higher than point $B_5$. There are also several crossings of $E_W$ vs $T$ in the right plot of Fig. \ref{fig:eopvsT1}, where different curves represent different values of $Q$. It is also easy to see the non-monotonic relationship between the $E_W$ and the $Q$.

In summary, the EWCS shows non-monotonicity along with charge $Q$ and the Lorentz violation parameter $\beta$ in Aether gravity. Specifically, the EWCS first decreases with the increase of them then increases with the increase of them. However, EWCS decreases monotonically with temperature $T$.

\section{Discussion}\label{sec:discuss}

In this paper, we studied the properties of mixed state entanglement and entanglement entropy in the Aether gravity theory with Lorentz symmetry violation. HEE and MI are found to monotonically change with the charge $Q$, the Lorentz violation parameter $\beta$, and temperature $T$. First, HEE increases with the increasing $Q$, $\beta$, $T$, and the width $w$. Moreover, as a measure of mixed state entanglement, MI shows exactly the opposite monotonicity to HEE. These results are independent of the specific configuration. More importantly, we found that EWCS behaves very differently from HEE or MI in Aether gravity. With the increasing $T$, EWCS decreases monotonically. However, EWCS shows non-monotonicity with $Q$ and $\beta$: it first decreases with the increasing $Q$ and $\beta$ and then increases with them. Based on the analytical treatments and numerical results, we show that EWCS behaves non-monotonically in the direction of $\beta$ due to the special role of Lorentz violation parameter $\beta$ in geometry. When the Lorentz violation parameter of the system is small, EWCS decreases rapidly with the increase of $\beta$. However, when the Lorentz violation parameter is large, that is, the $\beta$ is very large, it has little influence on the entanglement properties when further increasing $\beta$. These phenomena will lay a foundation for the further study of the gravity system with Lorentz violation.

One of the topics worthy of further study is to examine other gravity models with Lorentz violation to check whether the effect of Lorentz violation is consistent with those discussed in this paper. In addition, it is worth exploring other models that do not have explicit Lorentz violation but have obvious Lorentz violation in their dual systems. It is desirable to reveal the properties of mixed state entanglement in these models to further understand the influence of Lorentz violation on quantum information-related physical quantities. The above research will help to understand the relationship between systems with and without explicit Lorentz violation, and to further understand the duality of Lorentz violation in condensed matter theory.

\section*{Acknowledgments}

Peng Liu would like to thank Yun-Ha Zha for her kind encouragement during this work. This work is supported by the Natural Science Foundation of China under Grant No. 11805083, 11905083, 12005077 and Guangdong Basic and Applied Basic Research Foundation (2021A1515012374)

\end{document}